\documentclass[aps,prl,twocolumn,superscriptaddress,longbibliography]{revtex4-2}
\usepackage{amsmath}
\usepackage{graphicx}
\usepackage{hyperref}
\usepackage{color}
\usepackage{amsfonts}
\usepackage{amssymb}
\usepackage{braket}
\usepackage{times}
\usepackage{natbib}
\usepackage{lineno}
\usepackage{siunitx}

\begin{document}

\title{\huge \textbf{Coherent control of photon pairs via quantum interference between second- and third-order\\
quantum nonlinear processes} 
}%

\author{Alessia Stefano}
\email{Contact author: alessia.stefano01@universitadipavia.it} 
\affiliation{Dipartimento di Fisica, Università di Pavia, via A. Bassi 6, 27100 Pavia, Italy}

\author{Samuel E. Fontaine, J.E. Sipe}
\affiliation{
Department of Physics, University of Toronto, 60 St. George Street, Toronto, Ontario M5S 1A7, Canada
}

\author{Marco Liscidini}%
\affiliation{Dipartimento di Fisica, Università di Pavia, via A. Bassi 6, 27100 Pavia, Italy
}%

\begin{abstract}

Genuine quantum interference between independent nonlinear processes of different order provides a route to coherent control that cannot be reduced to a classical field interference. Here we present an all-optical analogue of coherent carrier injection by exploiting interference between second- and third-order quantum nonlinear processes in an integrated photonic platform. Photon pairs generated via spontaneous parametric down-conversion and spontaneous four-wave mixing coherently contribute to the same final two-photon state, resulting in a phase-dependent modulation of both the generation rate and the spectral structure of the emitted biphoton state. We illustrate the features of such interference and how it can be used to shape biphoton wavefunctions and their quantum correlations. These results identify interference between nonlinear processes of different order as a distinct form of coherent quantum control within quantum nonlinear optics.

\end{abstract}

\maketitle












Interference is a fundamental concept in physics, appearing both in the classical and quantum domains. A paradigmatic example is the double-slit experiment \cite{young1804experiments, born2013principles}. Within classical electromagnetism, it demonstrates the wave nature of light: the intensity observed on the screen results from the interference of the contributions passing through the two slits. When the experiment is performed with particles \cite{tonomura1989demonstration}, their wave-like character also emerges, but the situation is fundamentally different. The probability of detecting an electron at a given position originates from the superposition of quantum probability amplitudes associated with the two possible paths. In this case, the interference pattern is therefore a distinctive signature of a quantum effect. As interference is such a central element in both classical and quantum frameworks, it is not always straightforward to determine whether an interference pattern should be regarded as a genuinely quantum result, or whether it can still be described as the manifestation of a classical interference.
\begin{figure}[t]
    \centering
    \includegraphics[width=\linewidth]{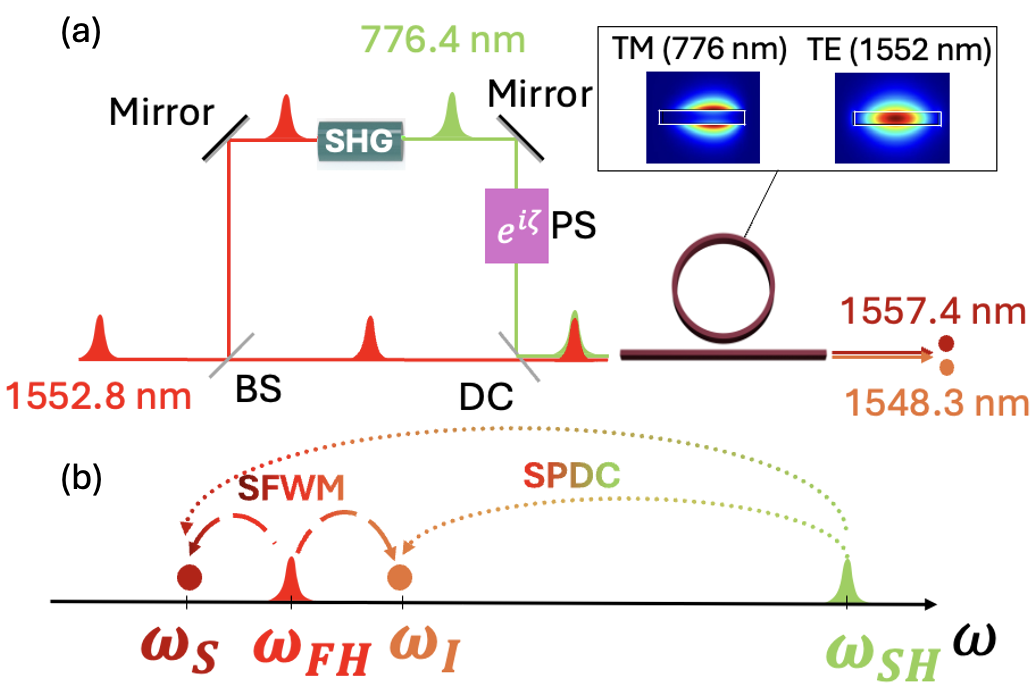}
    \caption{(a) Schematic of the proposed setup. BS: beam splitter; DC: dichroic mirror; PS: phase shifter. Inset: simulated TM and TE modes at $776.4$ nm and $1552.8$ nm, respectively. (b) Energy diagram of the nonlinear interactions. 
}
    \label{fig:1}
\end{figure}
In this regard, an instructive example is the experiment of Herzog et al. \cite{herzog1994frustrated}, in which interference is exploited to control the intensity of spontaneous parametric down-conversion (SPDC). In their setup, SPDC takes place in a single nonlinear crystal, where a pump beam propagates through the medium and generates photon pairs. Mirrors are then used to reflect both the pump and the down-converted fields, making photon pairs that would be produced in the forward and backward directions indistinguishable. By adjusting the relative phase of the counter-propagating pump fields, SPDC can be suppressed, a mechanism known as “frustrated two-photon creation.”
The authors interpreted this effect as the destructive interference of two quantum nonlinear processes, namely forward and backward SPDC. In this picture, SPDC suppression arises from the quantum interference of probability amplitudes associated with pair generation in each direction. 

An alternative view, however, is that, while SPDC itself is inherently a quantum process, the \textcolor{black}{observed interference can be understood as occurring at the classical level, \emph{before} quantization of the electromagnetic field. At each frequency, the forward-propagating and reflected fields combine to form a standing wave, which can then be used as a mode basis for quantization \cite{liscidini2012asymptotic}. The resulting Hamiltonian contains a spatial overlap integral of these standing-wave modes. By changing the positions of the standing-wave nodes and antinodes, the mirrors can cause the nonlinear spatial overlap integral to vanish.}
In this description, the SPDC suppression results from a classical interference of electromagnetic fields, without invoking the superposition of quantum probability amplitudes. Both perspectives are consistent and account for the observations, yet the existence of such dual interpretations raises the question of whether one is truly observing quantum interference of quantum processes, or rather a classical interference taking place in a quantum process. 




A genuine case of quantum interference between distinct processes is instead realized in coherently controlled injection currents in graphene \cite{atanasov1996coherent,sun2010coherent} and in thin films of topological insulator \cite{bas2015coherent}. In these experiments,  ballistic photocurrents are launched via the coherent control of two competing excitation pathways: one-photon absorption at $2\omega$ and two-photon absorption at $\omega$. 
\textcolor{black}{Because the two excitation pathways are driven by fields at different frequencies, they cannot interfere at the linear level of the classical electromagnetic fields; rather, the interference occurs \emph{after} quantization. Thus, the observed phase-dependent modulation of the expectation value of the generated current must arise from the coherent superposition of the probability amplitudes associated with the two distinct quantum processes. This conclusion is ultimately a consequence of the fact that the two pathways correspond to nonlinear processes of different order: one-photon absorption is linear in the pump intensity, whereas two-photon absorption is quadratic.}

\textcolor{black}{As illustrated by the two examples discussed above, the possibility of interpreting an interference effect either before or after field quantization is related to the structure of the interfering processes. In the experiment by Herzog \textit{et al.}, the two pathways are driven by fields at the same frequency. Since field quantization is defined in terms of the normal modes of the linear Maxwell equations, different choices of mode basis can change the number of modes effectively interacting in the considered SPDC configuration, thus leading to either two distinct terms or a single effective Hamiltonian. Consequently, the observed interference can admit both a classical interpretation, occurring before quantization, and a quantum interpretation, occurring after quantization. This ambiguity is no longer possible when the interfering quantum processes involve different numbers of annihilation operators. In this case, the pump fields cannot interfere at the classical level before quantization. The interference arise only because of the nonlinear quantum processes themselves, which coherently contribute to the same final state and interfere at the level of probability amplitudes.} This observation suggests that unambiguous quantum interference should be sought in situations where nonlinear processes of different order coherently contribute to the same final state.

In this work, we argue that an all-optical analogue of coherent current injection can occur in nonlinear photonic structures. In particular, we consider an integrated ring resonator where strong optical confinement enhances nonlinear interactions of different orders, enabling the generation of photon pairs with comparable rates and overlapping spectral bandwidths. Photon pairs at the same frequencies can be produced either through SPDC, a process involving a second-order response coefficient in the electric field, or through spontaneous four-wave mixing (SFWM), a process involving a third-order response coefficient in the electric field, with both processes pathways coherently contributing to the same final two-photon state. 

\textcolor{black}{Crucially, the two processes are described by Hamiltonians containing different numbers of annihilation operators ($\hat{H}_\mathrm{SPDC}\propto \hat{a}_{Sk_1}^\dagger \hat{a}_{Ik_2}^\dagger \hat{a}_{Pk}$ and $\hat{H}_\mathrm{SFWM}\propto \hat{a}^\dagger_{Sk_1}\, \hat{a}^\dagger_{Ik_2}\,\hat{a}_{Pk_3}\, \hat{a}_{Pk_4}$). \textcolor{black}{Unlike the experiment of \textcolor{black}{Herzog \textit{et al.},} 
here the two pumps have different frequencies, \textcolor{black}{and} thus interference \textcolor{black}{can} occur only at the level of probability amplitudes when the two distinct quantum processes, SPDC and SFWM, are considered. Each process is associated with a distinct quantum generation \textcolor{black}{pathway,} 
giving rise to phase-dependent interference of \textcolor{black}{a} genuinely quantum origin.} This interference can be exploited to directly modulate the joint spectral amplitude of the emitted photon pairs, providing a route to coherent control over their spectral correlations and generation rates.}

\begin{figure}[t]
    \centering
    \includegraphics[width=\linewidth]{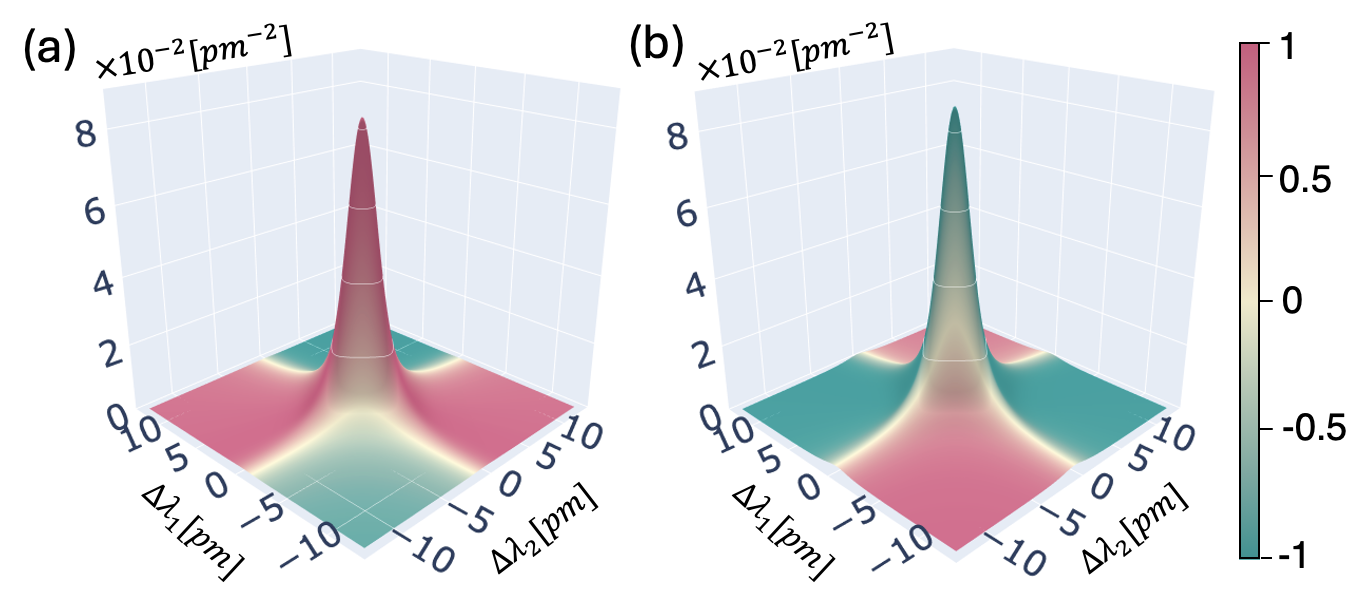}
    \caption{
    Plots of (a) $|\phi_\mathrm{SPDC}(\lambda_1,\lambda_2)|^2$[$pm^{-2}$] and (b) $|\phi_\mathrm{SFWM}(\lambda_1,\lambda_2)|^2$[$pm^{-2}$]. In the color, we encode the cosine of the phase.
    The SPDC pump has a pulse energy of $\SI{10.9}{pJ}$ while the SFWM pump has a pulse energy of $\SI{220}{pJ}$. Both pumps have a Gaussian spectral shape and a pulse duration of \SI{100}{ps}. }
    \label{fig:2}
\end{figure}

Here we consider SPDC and SFWM occurring within a high-Q integrated microring resonator, with photon pairs generated at telecom wavelengths around $1550$ nm. While several material platforms, such as AlGaAs \cite{baboux2023nonlinear,steiner2023continuous,meskine2025experimental} or LN \cite{lu2019periodically,lu2023generation,stefano2024broadband}, could be considered, in our example we focus on an In$_x$Ga$_{1-x}$P ($x=0.49$) microring, with second-order and third-order nonlinear susceptibilities of $\chi^{(2)} \simeq\SI{220}{pm/V}$ and $\chi^{(3)} \simeq 1.1\times 10^{-14} \mathrm{cm^2/V^2}$, respectively \cite{thiel2024wafer}. \textcolor{black}{The spatial overlap between this modes is taken into account in our calculations in the effective area, which is of the order of some 
$\SI{0.5}{\micro m^2}$ for both processes \cite{yang2008spontaneous, fontaine2025photon} (See Supplementary Materials)}. We assume the microring resonator to be fully embedded in SiO$_2$. 
\textcolor{black}{The device features a waveguide cross-section of width $w = \SI{1202}{nm}$ and height $h = \SI{102}{nm}$, which ensure phase matching for SFWM at around $\lambda_{\mathrm{F}} = 1552.8$ nm. Finally, we consider a radius of $R = \SI{33}{\micro\meter}$ to implement angular phase matching between the fundamental TE and TM modes operating at $\lambda_{\mathrm{F}} = 1552.8$ nm and $\lambda_{\mathrm{SH}} = 776.4$ nm, respectively.}
The dispersion relations of the fundamental modes of the waveguides are obtained from eigenmode simulations \cite{Lum_Mode} and expanded up to the second order around $\lambda_{\mathrm{F}}$ nm for the TE polarization, with $n_{\mathrm{eff,F}}=1.78$, $n_{\mathrm{g,F}}=2.50$, and \textcolor{black}{the group velocity dispersion is} $\beta_{2,\mathrm{F}}=2.70\times 10^{-24} \,\mathrm{s^2/m}$,  and around $\lambda_{\mathrm{SH}}$ nm for the TM polarization, with $n_{\mathrm{eff,SH}}= 1.69$, $n_{\mathrm{g,SH}}=2.47$, and $\beta_{2,\mathrm{SH}}=2.57\times10^{-24}\,\mathrm{s^2/m}$.  
We assume the microring to be critically coupled to the bus waveguide with intrinsic quality factors $Q_{\mathrm{int},\mathrm{F}} \simeq 10^{6}$ and $Q_{\mathrm{int},\mathrm{SH}} \simeq 10^{5}$, compatible with the current state of the art \cite{fontaine2025photon,thiel2024wafer}. 

We envision the configuration illustrated in Fig.~\ref{fig:1}, designed to implement both single-pump SFWM and non-degenerate type-I SPDC. A portion of a TE-polarized pump pulse centered at the resonant wavelength $\lambda_{\mathrm{F}}$ undergoes type-I second-harmonic generation (SHG) in a nonlinear crystal, producing a coherent TM-polarized pulse at $\lambda_{\mathrm{SH}}$, with relative phase $\zeta$ between the two pulses controlled via a phase shifter. Both pulses are then simultaneously coupled into the ring resonator through a bus waveguide. Under these conditions, photon pairs are generated via SFWM and SPDC on resonances symmetrically located with respect to that at $\lambda_{\mathrm{F}}$.


\begin{figure*}
    \centering
    \includegraphics[width=0.95\linewidth]{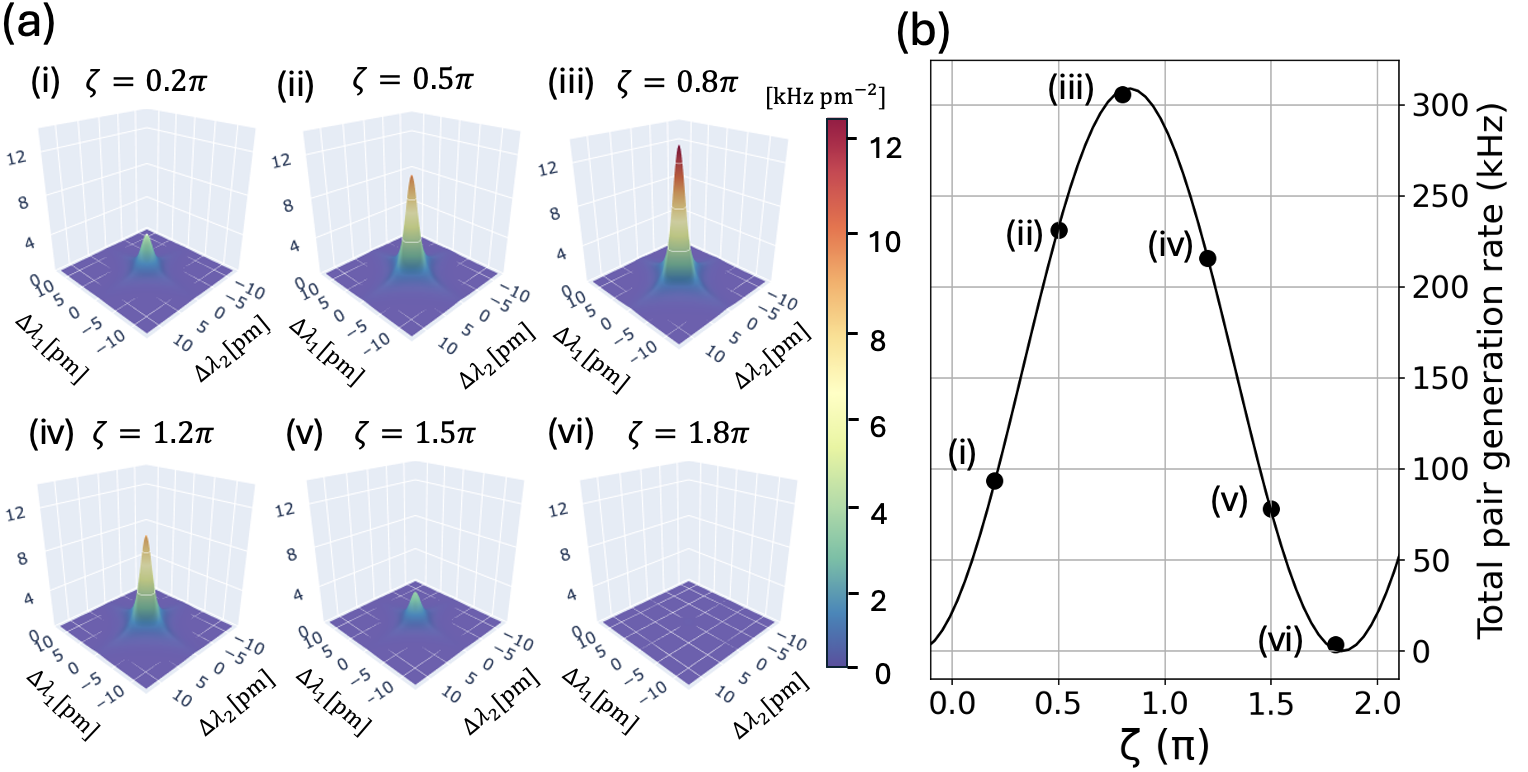}
    \caption{(a) Plots of $R_r|\beta_{\mathrm{TOT}}(\zeta)\phi_{\mathrm{TOT}}(\lambda_1,\lambda_2,\zeta) |^2 [\mathrm{kHz \,\,pm^{-2}}]$  for different relative phases $\zeta$. The parameters for SPDC and SFWM are the same as in Fig.~\ref{fig:2}. Quantum interference between the two processes modifies both the amplitude and the spectral correlations of the biphoton wavefunction.
    (b) Total pair generation rate given by Eq. \eqref{betatot}, for a repetition rate of 10 MHz. The dots correspond to the six cases shown in panel (a).}
    \label{fig:3}
\end{figure*}

In the limit of a low pair-generation probability, the output state of this system can be approximated by the normalized two-photon state
\begin{equation}
|\mathrm{II}\rangle=\dfrac{1}{\beta_{\mathrm{TOT}}}\left(\beta_{\mathrm{SFWM}}|\mathrm{II_{\mathrm{SFWM}}\rangle}+e^{i\zeta}\beta_{\mathrm{SPDC}}|\mathrm{II_{\mathrm{SPDC}}\rangle}\right) ,\label{II}
\end{equation}
where $|\mathrm{II_{\mathrm{SFWM (SPDC)}}}\rangle$ is the state that would be generated by SFWM (SPDC), $|\beta_{\mathrm{SFWM} (SPDC)}|^2$ is the corresponding average number of generated pairs per pulse, and $|\beta_{\mathrm{TOT}}|^2$ can be interpreted as the average number of photon pairs generated per pump pulse through the combination of SPDC and SFWM. Finally, $\zeta$ is the relative pump phase (see Fig.\ref{fig:1} (a)).

From the normalization of the two-photon state $|\mathrm{II}\rangle$ from Eq. \eqref{II} (i.e. $\langle \mathrm{II}|\mathrm{II}\rangle=1 $), and taking $|\mathrm{II}_{\mathrm{SPDC}}\rangle$ and $|\mathrm{II}_{\mathrm{SFWM}}\rangle$ to be normalized, it follows that
\begin{align}
    \label{betatot} |\beta_{\mathrm{TOT}}(\zeta)|^2=&|\beta_{\mathrm{SPDC}}|^2+|\beta_{\mathrm{SFWM}}|^2 +\\
2|\beta_{\mathrm{SPDC}}|&|\beta_{\mathrm{SFWM}}||\langle \mathrm{II}_{\mathrm{SFWM}}|\mathrm{II}_{\mathrm{SPDC}}\rangle|\cos(\zeta+\zeta_0)\nonumber ,
\end{align}
\textcolor{black}{where the complex amplitudes $\beta_{\mathrm{SFWM}}$ and $\beta_{\mathrm{SPDC}}$ each contain a process-dependent phase arising from the nonlinear coupling constants, propagation and coupling phases, and the phases of the driving pumps. The observable interference depends on the relative phase between these two amplitudes, which can be controlled through the phase shifter shown in Fig. 1. Thus, the phases of $\beta_{\mathrm{SPDC}}$, $\beta_{\mathrm{SFWM}}$, and $\langle \mathrm{II}_{\mathrm{SFWM}}|\mathrm{II}_{\mathrm{SPDC}}\rangle$ are absorbed into an overall offset phase $\zeta_0$, while $\zeta$ denotes the externally controlled phase governing the interference between the two pathways.} 
As usual, the term $|\langle \mathrm{II}_{\mathrm{SFWM}}|\mathrm{II}_{\mathrm{SPDC}}\rangle|$ can be interpreted as a measure of the indistiguishability of the states generated through SFWM and SPDC, where $|\langle \mathrm{II}_{\mathrm{SFWM}}|\mathrm{II}_{\mathrm{SPDC}}\rangle|\leq1$, with the equal sign holding when the two states are indistinguishable. One can also rewrite the total biphoton wavefunction as
\begin{align}
\label{phi_tot}
    \phi_{\mathrm{TOT}}(\omega_1,\omega_2,\zeta)=\dfrac{1}{\beta_{\mathrm{TOT}}} \bigg[&  \beta_{\mathrm{SFWM}} \phi_{\mathrm{SFWM}}(\omega_1,\omega_2)\nonumber\\
    +&e^{i\zeta} \beta_{\mathrm{SPDC}} \phi_{\mathrm{SPDC}}(\omega_1,\omega_2)\bigg],
\end{align}
where $\phi_{\mathrm{SPDC}}(\omega_1,\omega_2)$ and $\phi_{\mathrm{SFWM}}(\omega_1,\omega_2)$ are the biphoton wavefunction of the photon pairs that would be generated by SPDC and SFWM individually (see Supplementary Materials). In Fig~\ref{fig:2}(a) and Fig.~\ref{fig:2}(b) we show an example of such biphoton wavefunctions, where the surface height gives the $\phi_{\mathrm{SPDC}}(\omega_1,\omega_2)$ and $\phi_{\mathrm{SFWM}}(\omega_1,\omega_2)$ moduli, and the color map refers to the cosine of the phase of each biphoton wavefunction.

The results in Eqs.\eqref{betatot}  and \eqref{phi_tot} show that the generation rate and biphoton wavefunction of the generated pairs  can be controlled by exploiting the quantum interference of the individual processes. However, such an approach is effective only under two simultaneous conditions: (i) the pair-generation probabilities are comparable, i.e., $|\beta_{\mathrm{SFWM}}|^2 \sim |\beta_{\mathrm{SPDC}}|^2$, and (ii) the two processes generate photon pairs with similar spectral distributions. In bulk systems, fulfilling both requirements is extremely challenging, because second- and third-order nonlinear processes typically differ in efficiency significantly, and achieving phase matching over the same spectral range for both processes is difficult. Integrated photonic platforms can overcome these limitations by enabling SFWM and SPDC to coexist within the same ring resonator, which provides a stronger field enhancement for SFWM \cite{sharping2006generation,clemmen2009continuous,steiner2021ultrabright} and confines pair generation to the same frequency ranges set by the resonances. For example, using the pump parameters of Fig.~\ref{fig:2}, one can achieve generation rates of \SI{77}{kHz} for both SFWM and SPDC at a pump repetition rate $R_R$ of \SI{10}{MHz}, thereby satisfying the condition of comparable generation probabilities in addition to their spectrally overlapping emission.

Building on these results, we now show the effect of interference between the two biphoton wavefunctions as the relative pump phase $\zeta$ is varied. We first consider its impact on the generation rate. In Fig.~\ref{fig:3}(a), we plot the photon-pair spectral generation rate $|\beta_{\mathrm{TOT}}\phi_{\mathrm{TOT}}(\lambda_1,\lambda_2)|^2 R_r$ for increasing values of $\zeta$. By integrating over $\lambda_1$ and $\lambda_2$, we obtain the total generation rate for each configuration. For specific values of $\zeta$, destructive interference strongly suppresses pair generation (e.g., Fig.~\ref{fig:3}(vi)), while for others, constructive interference between the two processes leads to a substantial enhancement of the generation rate (e.g., Fig.~\ref{fig:3}a (iii)). \textcolor{black}{Here we focus on a specific resonance pair, while we assume pairs generated on  other modes of the resonator to be spectrally filtered.}

The total normalization factor $\beta_{\mathrm{TOT}}$, computed from the individually normalized biphoton wavefunctions, is shown in Fig.~\ref{fig:3}(b) together with the analytical expression in Eq.~\eqref{betatot}. The excellent agreement confirms that interference between SPDC and SFWM results in a phase-dependent enhancement or suppression of the overall photon-pair generation rate.

The spectral overlap $|\langle \mathrm{II}_{\mathrm{SFWM}}|\mathrm{II}_{\mathrm{SPDC}}\rangle|$ directly determines the interference visibility,
\begin{equation}
V = \frac{2|\beta_{\mathrm{SPDC}}\beta_{\mathrm{SFWM}}\langle \mathrm{II}_{\mathrm{SFWM}}|\mathrm{II}_{\mathrm{SPDC}}\rangle|}{|\beta_{\mathrm{SPDC}}|^2 + |\beta_{\mathrm{SFWM}}|^2}.
\end{equation}
Maximum visibility is obtained when the two wavefunctions have identical spectral profiles ($|\langle \mathrm{II}_{\mathrm{SFWM}}|\mathrm{II}_{\mathrm{SPDC}}\rangle| = 1$) and comparable generation probabilities ($|\beta_{\mathrm{SPDC}}| = |\beta_{\mathrm{SFWM}}|$). Any mismatch in spectral overlap or generation probability reduces the visibility. In particular, when the generation probabilities are equal, the visibility simplifies to $V = |\langle \mathrm{II}_{\mathrm{SFWM}}|\mathrm{II}_{\mathrm{SPDC}}\rangle|$, so the visibility directly quantifies the indistinguishability of the two biphoton states. In our resonant structure, the resonance-induced Lorentzian response enforces identical spectral shapes for the two biphoton wavefunctions, yielding an exceptionally high estimated visibility of $V = 0.99997$. This demonstrates that the interference is nearly ideal and that the two biphoton states are almost perfectly indistinguishable.

\begin{figure}
    \centering
    \includegraphics[width=0.7\linewidth]{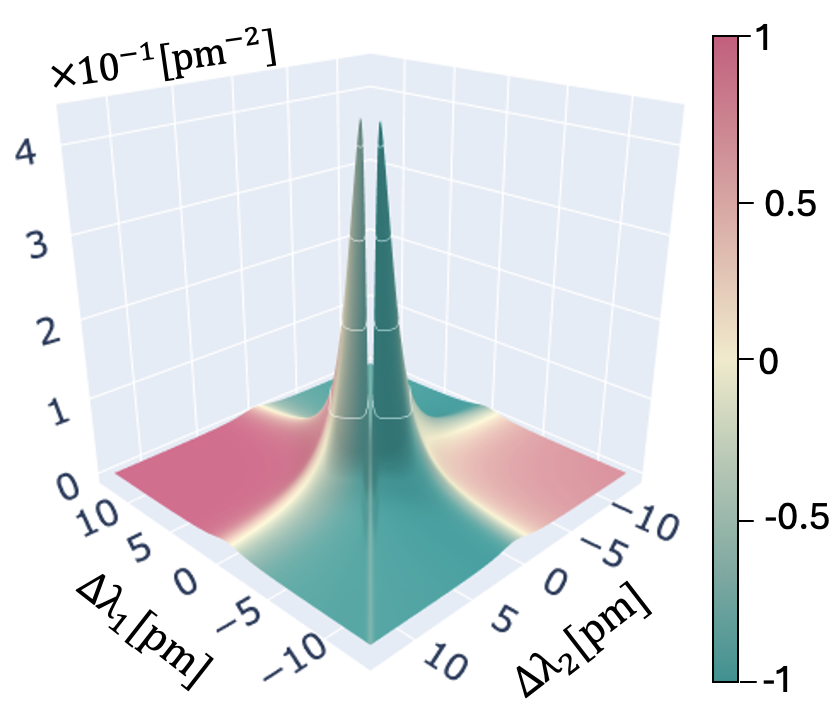}
    \caption{Plot of $|\phi_\mathrm{TOT}(\lambda_1,\lambda_2)|^2$ [pm$^{-2}$]. Here $\zeta=1.8\pi$, pulse duration of the SPDC pump is $\SI{3}{ns}$ and the one of SFWM is $\SI{100}{ps}$.}
    \label{fig:TotalBWF18pi}
\end{figure}

We now turn to the last aspect enabled by quantum nonlinear interference, namely the possibility of engineering highly structured biphoton wavefunctions. Beyond controlling the overall generation rate, interference between distinct nonlinear pathways provides a powerful tool to tailor the spectral correlations of the emitted photon pairs. Here we present a representative example illustrating this capability, while additional configurations are discussed in the Supplementary Material. To this end, we keep the SFWM parameters unchanged and modify only the temporal profile of the SPDC pump, which is taken to have a duration of $\SI{3}{ns}$ and a pulse energy of $\SI{0.5}{pJ}$. This choice significantly reshapes the corresponding SPDC biphoton wavefunction, producing a strongly elongated spectral lobe. The resulting SPDC wavefunction for the longer pump pulse is reported in the Supplementary Material. Because of the  markedly different spectral shapes of the SPDC and SFWM contributions, the interference is no longer uniform across the spectrum, as shown in Fig.\ref{fig:TotalBWF18pi}. Instead, it becomes highly localized, giving rise to a deep spectral notch only in the region where the two biphoton wavefunctions overlap. As a result, the total biphoton wavefunction is effectively split into two distinct spectral lobes separated by a pronounced trench. More generally, quantum interference between nonlinear processes can be exploited to synthesize biphoton wavefunctions with nontrivial structures that would be quite difficult (or even impossible) to obtain by a pump engineering in a simple geometry as the ring one considered here.

In conclusion, we have shown that genuine quantum interference between nonlinear processes of different order can be realized and harnessed in an integrated photonic platform. By coherently combining second- and third-order nonlinear pathways, we implement an optical analogue of coherent carrier injection, where interference arises from the superposition of fundamentally distinct quantum generation mechanisms.
Crucially, this interference leads to outcomes that are far richer than those encountered in the solid-state analogue. Instead of generating a macroscopic current, here interference is exploited to directly synthesize biphoton wavefunctions with tailored spectral correlations, enabling control over genuinely quantum properties such as entanglement. The resulting features would be difficult, if not impossible, to obtain in simple resonant systems relying on a single nonlinear process. \textcolor{black}{More importantly, our aim here is to clarify the nature of the interference underlying the generation of photon pairs when combining parametric processes of different orders.} Our work establishes quantum interference between nonlinear processes of different order as a distinct regime of coherent control, with implications that extend beyond quantum photonics. For example, this effect could be exploited to dynamically shape the parametric gain and thereby control the turn-on and pulsed dynamics of an optical parametric oscillator, providing a gain-side analogue of Q-switching in lasers. By enabling interference between fundamentally different quantum mechanisms, this study opens new directions for investigating and exploiting coherent phenomena across diverse areas of nonlinear and quantum physics.\\

\begin{acknowledgments}

A.S. and M.L. acknowledge PNRR MUR project “National Quantum Science and Technology Institute” – NQSTI (Grant No. PE0000023). 
S.E.F. and J.E.S. acknowledge the Natural Sciences and Engineering Research Council of Canada, and the European Union's Horizon Europe Research and Innovation Programme (101070700, project MIRAQLS) for financial support. S.E.F. acknowledges support from a Walter C. Sumner Memorial Fellowship.
\end{acknowledgments}


\bibliography{apssamp}

\end{document}


\title{\textbf{Supplementary Information for ``Coherent control of photon pairs via quantum interference of quantum nonlinear processes"} }%

\author{Alessia Stefano}\email{alessia.stefano01@universitadipavia.it}
\affiliation{Dipartimento di Fisica, Università di Pavia, via A. Bassi 6, 27100 Pavia, Italy}

\author{Samuel E. Fontaine}
\affiliation{Department of Physics, University of Toronto, 60 St. George Street, Toronto, Ontario M5S 1A7, Canada}

\author{J. E. Sipe}
\affiliation{Department of Physics, University of Toronto, 60 St. George Street, Toronto, Ontario M5S 1A7, Canada}

\author{Marco Liscidini}
\affiliation{Dipartimento di Fisica, Università di Pavia, via A. Bassi 6, 27100 Pavia, Italy}

\maketitle

\section{Biphoton wavefunctions calculation}
The {\color{black}Hamiltonian} 
$H_{{\color{black}\mathrm{NL}}}^{(2)}$ \textcolor{black}{for the second-order nonlinear processes} can be written as \cite{quesada2022beyond}
\begin{equation}
    H_{\mathrm{NL}}^{(2)}
    =-\dfrac{1}{3\epsilon_0}\int d \mathbf{r} \Gamma_2^{ijl}(\mathbf{r}) D^i(\mathbf{r})D^j(\mathbf{r})D^l(\mathbf{r}) \label{HSPDC}
\end{equation}
where $D^k(\mathbf{r})$ are the Cartesian components of the displacement field operators, and where $\Gamma_2$ is a nonlinear coefficient related to the second-order nonlinear susceptibility $\chi_2$ \cite{quesada2022beyond}

\begin{equation}
\Gamma^{ijk}_2(\mathbf{r})=\dfrac{\chi_2^{ijk}(\mathbf{r})}{\epsilon_0 n^2(\mathbf{r},\omega_{S,k_1})n^2(\mathbf{r},\omega_{I,k_2})n^2(\mathbf{r},\omega_{P,k_3})}{\color{black},}
\end{equation}
where $n(\mathbf{r},\omega)$ is the refractive index at position $\mathbf{r}$ and frequency $\omega$.


We consider spontaneous parametric down{\color{black}-}conversion (SPDC)\textcolor{black}{,} in which a pump photon at frequency $\omega_{P,k_3}$ 
is
converted into two photons at frequency $\omega_{S,k_1}$ and $\omega_{I,k_2}$, labelled by {\color{black}$S$} (signal) and {\color{black}$I$} (idler) respectively. We can
write the Hamiltonian containing the energy conserving terms relevant to SPDC in terms of asymptotic fields \cite{liscidini2012asymptotic},
such that we have



\begin{align}
    H_{\mathrm{SPDC}} =& \int \mathrm{d}k_1\, \mathrm{d}k_2\, \mathrm{d}k_3 J_{\mathrm{SPDC}}(k_1,k_2,k_3)\, c_{Sk_1}^\dagger c_{Ik_2}^\dagger c_{Pk_3}+{\color{black}\mathrm{H.c.}},
\end{align}
where the operators have the usual bosonic commutation relations $[c_{mk},c^{\dagger}_{m' k'}]=\delta_{m m'}\delta(k-k')$. The nonlinear overlap
integral $J_\mathrm{SPDC}$ is written in terms of the asymptotic-in and -out fields of the structure of interest \cite{banic2022two,fontaine2025photon}
\begin{equation}
J_{\mathrm{SPDC}}(k_1,k_2,k_3) =
-\frac{2}{\varepsilon_0}
\int d\mathbf{r}\,
\Gamma^{ijl}_{2}(\mathbf{r})
[D^{\mathrm{out},{\color{black}i}}_{S,k_1}(\mathbf{r})
D^{\mathrm{out},{\color{black}j}}_{I,k_2}(\mathbf{r})]^{*}
D^{\mathrm{in},{\color{black}l}}_{P,k_3}(\mathbf{r}),
\end{equation}
and depends on the phase-matching condition of the given process. For \textcolor{black}{the} angular phase matching that appears in
microrings of III-V semi-conductors such as AlGaAs, the expressions can be found in 
\textcolor{black}{earlier work} \cite{liscidini2012asymptotic,fontaine2025photon}.



From the Hamiltonian, we can derive the biphoton wavefunction (BWF) \cite{banic2022two,fontaine2025photon}

\begin{align}
    \phi_{\mathrm{SPDC}}(k_1,k_2)
= \frac{2\pi\alpha_1}{\beta_{\mathrm{SPDC}}}\, \dfrac{i}{\hbar}\,
\int \mathrm{d}k_3\, \phi_P(k_3)\,
J_\mathrm{SPDC}(k_1,k_2,k_3)\,
\delta(\omega_{P,k_3}-\omega_{S,k_1}-\omega_{I,k_2}){\color{black},}
\end{align}
where $|\alpha_1|^2$ is the number of photons in the pump pulse relevant to the SPDC process, and $\phi_{P}(\omega) $ is a function
describing the spectral profile of the pump, assumed to be Gaussian in this work. 

To simplify integration, we move from a wavevector notation to a frequency notation, where we define the frequency
dependant quantities
\begin{equation}
    c_{m\omega} \equiv \sqrt{\frac{1}{v_{g,m}(\omega)}}\, c_{mk}
\end{equation}
\begin{equation}
\phi_P(\omega) \equiv  \sqrt{\frac{1}{v_{g,P}(\omega)}}\, \phi_P(k)
\end{equation}
\begin{equation}
\phi(\omega_1,\omega_2)
\equiv  \sqrt{\frac{1}{v_{g,S}(\omega_1)\, v_{g,I}(\omega_2)}}\, 
\phi(k_1,k_2)
\end{equation}
where  $v_{g,m}(\omega)=\dfrac{d\omega_m}{dk_m}$ is the group velocity of the mode {\color{black}$m$, and the operators have the usual bosonic commutation relations $[a_{m\omega},a^\dagger_{m'\omega'}]=\delta_{mm'}\delta(\omega-\omega')$}.
After integrating over $\omega_3$ with the Dirac delta function, we can write the BWF as
\begin{align}
    \phi_{\mathrm{SPDC}}(\omega_1,\omega_2)
=\frac{2\pi\alpha_{\color{black}1}}{\beta_{\mathrm{SPDC}}}\, \dfrac{i}{\hbar}\,
{\color{black}\sqrt{\frac{1}{v_{g,S}(\omega_1)\, v_{g,I}(\omega_2)v_{g,P}(\omega_1+\omega_2)}}} 
\,
\phi_P(\omega_1+\omega_2)\,
J_\mathrm{SPDC}(\omega_1,\omega_2,\omega_1+\omega_2)
\end{align}
where the probability of generating a pair $|\beta_{\mathrm{SPDC}}|^2$ is fixed by the normalization condition of the BWF
\begin{equation}
    \int d\omega_1 d\omega_2| \phi_{\mathrm{SPDC}}(\omega_1,\omega_2)|^2=1.
\end{equation}


The Hamiltonian 
$H_\mathrm{NL}^{(3)}$ \textcolor{black}{for third-order nonlinear processes} {\color{black}can be written as} \cite{quesada2022beyond}
\begin{equation}
    H_{\mathrm{NL}}^{(3)}
    =-\dfrac{1}{4\epsilon_0}\int d\mathbf{r} \Gamma_3^{ijlm}(\mathbf{r}) D^i(\mathbf{r})D^j(\mathbf{r})D^l(\mathbf{r})D^m(\mathbf{r}),
\end{equation}
where $\Gamma_3$ is related to the second- and third-order nonlinear susceptibilities{\color{black}} 
\cite{quesada2022beyond}.
Similar to the SPDC case, we can write
the nonlinear Hamiltonian containing the energy conserving terms relevant for single-pump spontaneous four-wave
mixing (SFWM) as
\begin{align}
    H_{\mathrm{SFWM}} = 
\int \mathrm{d}k_1\, \mathrm{d}k_2\, \mathrm{d}k_3\, \mathrm{d}k_4\,
J_{\mathrm{SFWM}}(k_1,k_2,k_3,k_4)\,
c^\dagger_{Sk_1}\, c^\dagger_{Ik_2}\,c_{Pk_3}\, c_{Pk_4}+\mathrm{H.c.}.\nonumber
\end{align}
This Hamiltonian describes the annihilation of two pump photons at frequencies $\omega_{P,k_3}$ and $\omega_{P,k_4}$, and the creation of
signal and idler photons at frequencies $\omega_{S,k_1}$ and $\omega_{I,k_2}$ respectively. The nonlinear overlap integral $J_\mathrm{SFWM}$ written
in terms of asymptotic fields for SFWM is given by


\begin{align}
    J_\mathrm{SFWM}(k_1,k_2,k_3,k_4)=-\frac{3}{\epsilon_0}\int d\mathbf{r}\,\Gamma_{3}^{ijlm}(r) 
\left[D^{\mathrm{out}{\color{black},i}}_{S,k_1}(\mathbf{r})\,
D^{\mathrm{out}{\color{black},j}}_{I,k_2}(\mathbf{r})\right]^*\,
D^{\mathrm{in}{\color{black},l}}_{P,k_3}(\mathbf{r})\,
D^{\mathrm{in}{\color{black},m}}_{P,k_4}(\mathbf{r}),
\end{align}
\textcolor{black}{see, e.g., Banic et al.} 
\cite{banic2022two}. 

Similar to SPDC, we can obtain an expression for the BWF associated to SFWM in the frequency domain

\begin{align}
    \phi_{\mathrm{SFWM}}(\omega_1,\omega_2)
=
\frac{2\pi(\alpha_2)^2} 
{\beta_\mathrm{SFWM}}\, \dfrac{i}{\hbar}\,
\sqrt{\frac{1}{v_{g,S}(\omega_1)\, v_{g,I}(\omega_2)}}
\int \mathrm{d}\omega_3\,
\phi_P(\omega_1+\omega_2-\omega_3)\,
\phi_P(\omega_3)\,
\frac{
J_\mathrm{SFWM}(\omega_1,\omega_2,\omega_3,\omega_1+\omega_2-\omega_3)
}{
\sqrt{v_{g,P}(\omega_3)\, v_{g,P}(\omega_1+\omega_2-\omega_3)}
}
\end{align}
where $ |\alpha_2|^2 $ is the number of photons in the pump pulse, and $|\beta_\mathrm{SFWM}|^2$ is the probability of generating a pair of signal
and idler photons from the SFWM process, found by normalizing the BWF

\begin{equation}
    \int d\omega_1 d\omega_2| \phi_{\mathrm{SFWM}}(\omega_1,\omega_2)|^2=1.
\end{equation}

To calculate $|\beta_\mathrm{SPDC}|^2$ and $|\beta_\mathrm{SFWM}|^2$ we compute the overlap integrals which contain the various fields involved in the
processes. These also depend on the structure of interest and phase-matching conditions, and explicit expressions for
these quantities can be found \textcolor{black}{in earlier work} 
\cite{liscidini2012asymptotic,banic2022two, fontaine2025photon,zatti2023generation, yang2008spontaneous}.

\begin{figure}[h!]
    \centering
    \includegraphics[width=0.5\columnwidth]{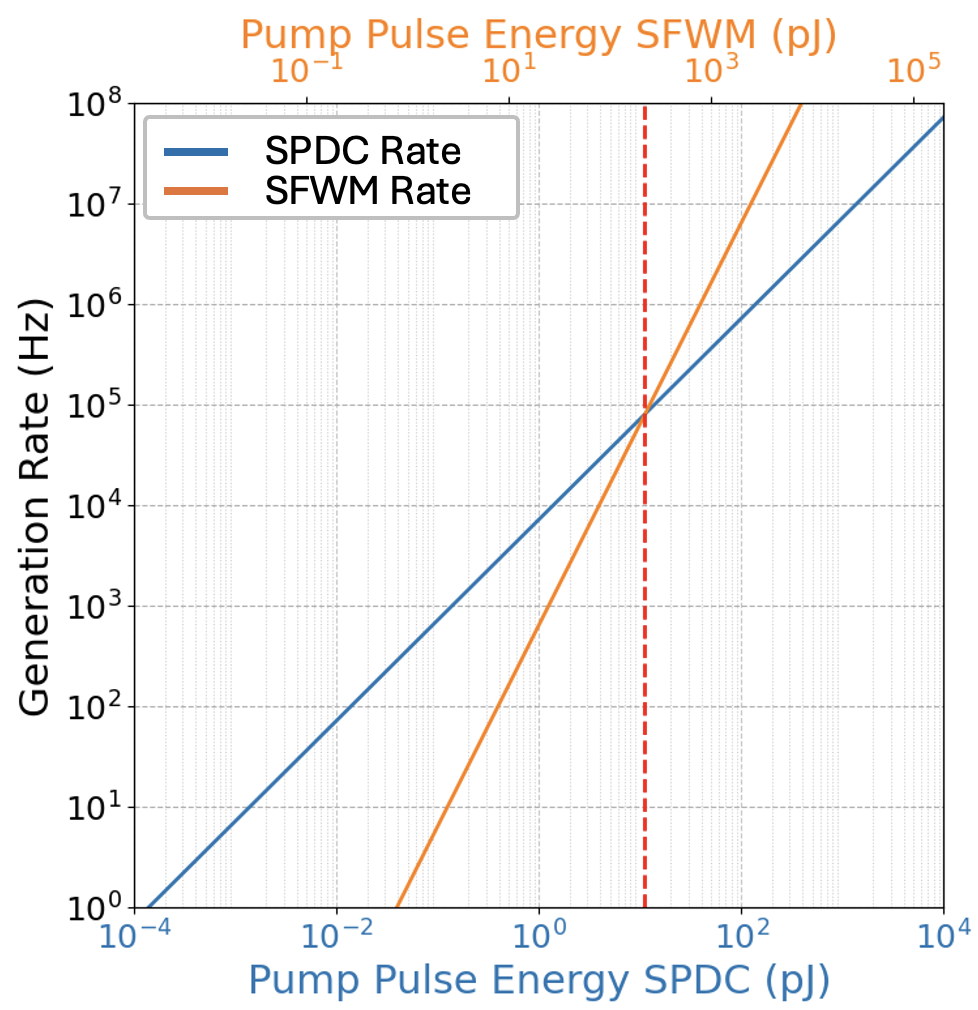}
    \caption{Calculated SPDC and SFWM pair-generation rates on a double-logarithmic scale as a function of pump pulse energy. The top axis refers to the SFWM pumps, the bottom to the SPDC pump. The SPDC curve shows a slope of 1 (linear 
    {\color{black}scaling}), while the SFWM curve shows a slope of 2 (quadratic scaling). For suitable pump energies, the two processes reach comparable efficiencies.}

    \label{fig:rates}
\end{figure}
Fig{\color{black}.}~\ref{fig:rates} shows the calculated SPDC and SFWM photon-pair generation rates on a log–log scale as a function of the
pump pulse energies, and we show the intersection where both generation rates are the same. The lower blue (upper
orange) $x$-axis refers to the energy of the pump pulse which excites the SPDC (SFWM) process. As expected, SPDC
scales linearly with pump energy, whereas SFWM scales quadratically.

\section{Long-Pulse SPDC and Spectral-Correlation Reshaping}

\begin{figure}[h!]
    \centering
    \includegraphics[width=0.5\columnwidth]{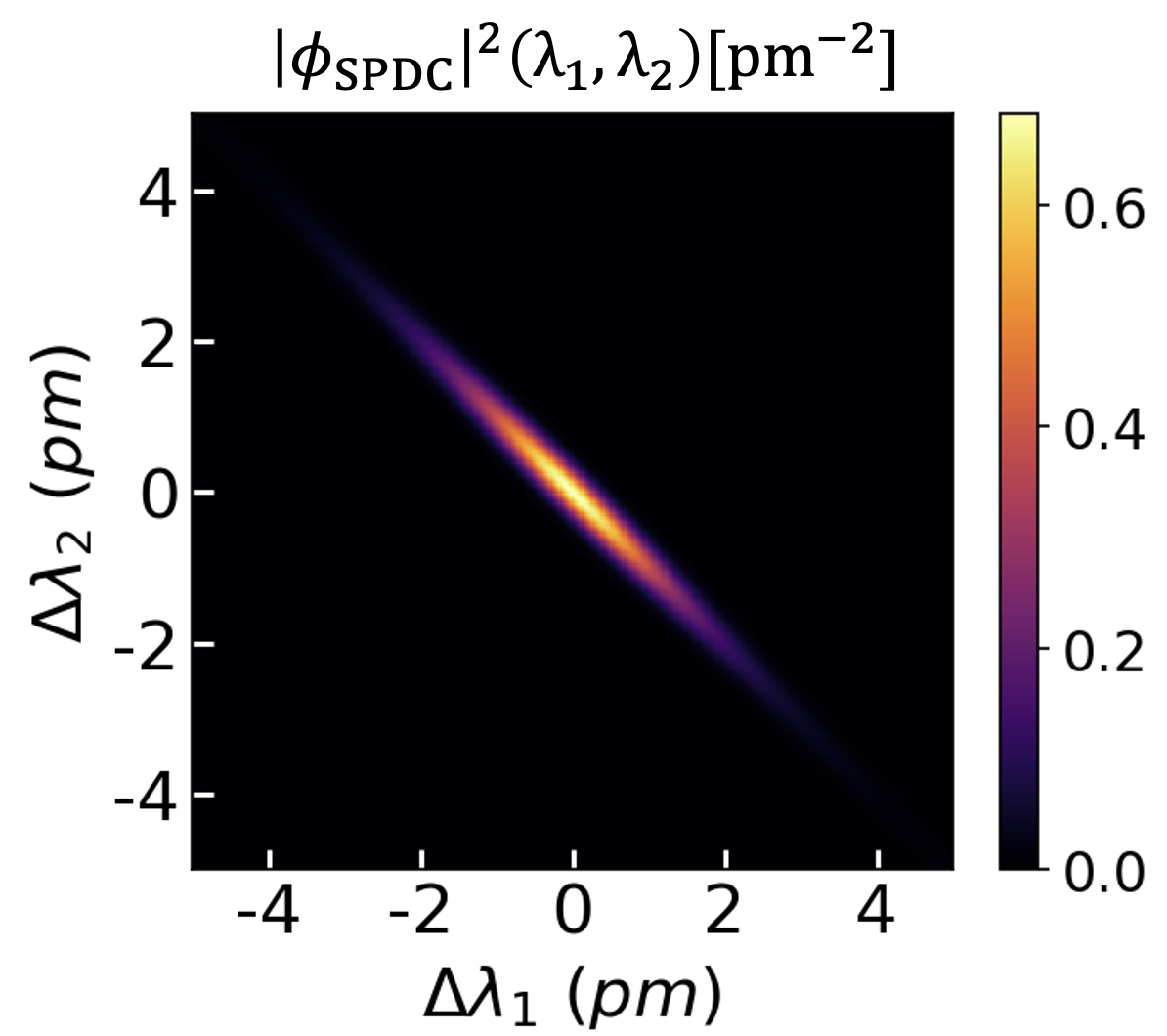}
    \caption{$|\phi_\mathrm{SPDC}|^2$ of the SPDC process for a \SI{3}{ns} pump pulse. The long pulse produces a narrow, elongated spectrum with limited overlap with the $|\phi_\mathrm{SFWM}|^2$ of SFWM process.}
    \label{fig:SPDC3ns}
\end{figure}

We present the effects that elongating the pump pulse for SPDC to \SI{3}{ns} have on the BWF, which is relevant for
the interference scenario shown in Fig. 4 in the main text.

Fig{\color{black}.}
~\ref{fig:SPDC3ns} shows the 
{\color{black}square-modulus of the BWF for} the SPDC process {\color{black}$|\phi_\mathrm{SPDC}|^2$} under this long-pulse excitation. 
As expected from the reduced pump bandwidth, the spectrum becomes significantly narrower along the energy-conservation axis, yielding a highly elongated and sharply peaked structure. 
\textcolor{black}{And as} {\color{black}
expected, $|\phi_{\mathrm{SPDC}}|^2$ generated with a 3 ns \textcolor{black}{pump} pulse is markedly different than $|\phi_{\mathrm{SFWM}}|^2$ generated with a 100 ps \textcolor{black}{pump} pulse, showcasing the mismatch that causes the lack of spectral overlap between the two processes. }

\bibliographystyle{apsrev4-2}
\bibliography{apssamp}